\documentstyle[preprint,eqsecnum,aps]{revtex}

\begin{document}
\draft
\preprint{}
\title{Incommensurate Magnetic Fluctuations in YBa$_2$Cu$_3$O$_{6.6}$}
\author{Pengcheng Dai,$^1$ H. A. Mook,$^1$ and F. Do$\rm\breve{g}$an$^2$}
\address{
$^1$Oak Ridge National Laboratory, Oak Ridge, 
Tennessee 37831-6393\\
$^2$Department of Materials Science and Engineering\\ 
University of Washington, Seattle, Washington 98195\\}
\date{\today}
\maketitle
\begin{abstract}
We use inelastic neutron scattering to demonstrate that at 
low temperatures, the 
low frequency 
magnetic fluctuations in YBa$_2$Cu$_3$O$_{6.6}$ ($T_c=62.7$ K) are 
incommensurate, being found at positions displaced by $\pm\delta$ ($0.057\pm0.006$ r.l.u.) along the 
$[\pi,\pi]$ direction from the wave vector $(\pi,
\pi)$ associated 
with the antiferromagnetic order of the parent insulator, YBa$_2$Cu$_3$O$_{6}$. 
The dynamical susceptibility $\chi^{\prime\prime}({\bf q},\omega)$ at
the incommensurate positions increases on cooling below $T_c$,
accompanied by a suppression of magnetic fluctuations at the commensurate points.
\end{abstract}

\narrowtext
Knowledge of the  
 spin dynamical properties of the cuprates are crucial to the understanding of
high-temperature ($T_c$) superconductivity.
An important issue is the symmetry of 
the imaginary part of the dynamical 
susceptibility, $\chi^{\prime\prime}({\bf q},\omega)$,
probed directly by neutron scattering.
Over the past several years,
 intensive experimental work on the single layer La$_{2-x}$Sr$_x$CuO$_4$ (214) 
\cite{cheong,yamada} and the 
bilayer YBa$_2$Cu$_3$O$_{7-x}$ [(123)O$_{7-x}$] 
\cite{mignod,tranquada,mook,fong,bourges,dai1,fong1} cuprates has yielded 
valuable information concerning the magnetic response in the normal and superconducting
states. For the 214 family, magnetic fluctuations were
found at incommensurate positions from 
the antiferromagnetic ordering position $(\pi,\pi)$ \cite{cheong}. 
For (123)O$_{7-x}$, the situation is more subtle. While Rossat-Mignod and 
coworkers only detected magnetic fluctuations at the commensurate position
$(\pi,\pi)$ \cite{mignod}, Tranquada {\it et al.} 
\cite{tranquada} noticed that 
the wave vector dependence of $\chi^{\prime\prime}({\bf q},\omega)$  
in a (123)O$_{6.6}$ ($T_c=53$ K) crystal was better
described by a pair of identical Gaussians displaced symmetrically from the
$(\pi,\pi)$ point, suggesting possible incommensurate 
fluctuations. In a
subsequent experiment, Sternlieb {\it et al.} \cite{tranquada} have
suggested that the wave vector 
dependence of the susceptibility is independent of energy
for all energies between 2 mV and 40 meV. Although the 
nontrivial wave vector dependence behavior  
 was again observed, $\chi^{\prime\prime}({\bf q},\omega)$ was not found to be 
different below and above $T_c$.  No firm conclusions about the commensurability and symmetry
of $\chi^{\prime\prime}({\bf q},\omega)$ were reached \cite{tranquada}.

In this Letter, we 
present inelastic neutron scattering data which 
resolves the issue of commensuration in (123)O$_{6.6}$. We show that 
the magnetic response is complex with incommensurate fluctuations for energies below
the commensurate resonance \cite{dai1} at low temperatures. 
The low frequency spin fluctuations change from commensurate to incommensurate
on cooling with 
the incommensuration first appearing at temperatures somewhat above $T_c$. 
Lowering the temperature
 suppresses the spin fluctuations at the commensurate points,
accompanied by an increase in the susceptibility at the incommensurate positions.

Our results shed light on several long standing theoretical predictions.  
If the (123)O$_{7-x}$ system is indeed a $d$-wave superconductor,  $d$-wave 
gap nodes could 
yield incommensurate peaks below $T_c$ \cite{lu}. In this scenario, 
the scattering of (123)O$_{6.6}$ is expected 
to be commensurate in the normal state and incommensurate 
in the superconducting state \cite{zha}.  
No neutron scattering experiment, until now, has observed this  behavior.
Second, even though the differences in the commensurability in 
(123)O$_{7-x}$ and 
214 can be explained from the Fermi surface differences between the two compounds \cite{si}, 
 such models have not been entirely reconciled with the nuclear magnetic 
resonance (NMR) measurements \cite{walsted,barzykin}.  
The reconciliation of the neutron and NMR results requires detailed information about 
the structure of $\chi^{\prime\prime}({\bf q},\omega)$ \cite{zha1}. 
The third theoretical conjecture stems from recent 
neutron scattering experiments of Tranquada {\it et al.} \cite{tranquada1} which
suggest that the incommensuration in superconducting 
214 may be associated with a 
spatial segregation of 
charge or charge density wave correlations \cite{kivelson}. If the idea of 
dynamical microphase
separation in the CuO$_2$ plane asserted by Emery and Kivelson \cite{emery} 
is relevant for the high-$T_c$ superconductivity, 
one would expect incommensurate spin fluctuations in other cuprate
superconductors like Bi$_2$Sr$_2$CaCu$_2$O$_{8-x}$ (2212) and (123)O$_{7-x}$.  
Indeed, recent
neutron scattering experiments by Mook and Chakoumakos \cite{mookchako} 
 show that the spin fluctuations in 2212 
are also incommensurate. Thus, incommensurability may be a common feature
for all cuprate superconductors.

The neutron scattering
 measurements were made at the High-Flux Isotope Reactor at
Oak Ridge National Laboratory  
using the HB-1 and HB-3 triple-axis 
spectrometers. The fabrication and characteristics of our single-crystal 
sample of (123)O$_{6.6}$ (weight 25.59 grams and $T_c=62.7$ K) 
were described in detail  
previously \cite{dai1}. The major difficulty in studying spin fluctuations in 
the (123)O$_{7-x}$ system is to separate the magnetic scattering from (single- and multi-)
phonon and other 
spurious processes.  
While multi-phonon scattering usually has 
a simple wave vector dependence and spurious events 
such as accidental Bragg scattering can be identified by
checking the desired inelastic scan in the two-axis mode \cite{tranquada}, two approaches 
can be used to separate magnetic from single-phonon scattering.
The first approach is to perform neutron polarization analysis \cite{moon} which,
in principle, allows an unambiguous separation of magnetic and nuclear scattering. 
This method has been successfully employed to identify 
the magnetic origin of 
resonance peaks for ideally \cite{mook} and underdoped \cite{dai1,fong1} (123)O$_{7-x}$.
However, this advantage comes at a considerable cost in intensity which makes
the technique impractical for observing small magnetic signals. 
The second approach is to utilize the differences in 
the temperature and wave vector dependence of the phonon 
and magnetic scattering cross sections.
While phonon scattering gains intensity on warming due to the 
thermal population factor, the 
 magnetic signal usually becomes weaker  
because it spreads throughout the 
energy and momentum space at high temperatures. 
Thus, in an unpolarized neutron measurement the net 
intensity gain above the multi-phonon background
 on cooling at appropriate wave vectors 
is likely to be magnetic in origin.

Figure 1(a) depicts the reciprocal 
space probed in the experiment with {\bf a*} (=$1/a$), {\bf b*} (=$1/b$) 
directions shown in the square lattice notation. The momentum transfers 
$(q_x,q_y,q_z)$ in units of \AA$^{-1}$ are at positions 
$(H,K,L)=(q_xa/2\pi,q_yb/2\pi,q_zc/2\pi)$ reciprocal lattice units (r.l.u.). 
We first describe measurements made 
in the $(H,H,L)$ zone. Our search for the magnetic fluctuations was done with 
the filter integration technique \cite{mook2} first developed to study the chain fluctuations
in (123)O$_{6.93}$. This technique is excellent for isolating scattering from lower 
dimensional objects and relies on integrating the energy along wave vector
direction $[0,0,L]$ perpendicular to the scan direction $[H,H,0]$. 
 To estimate the energy integration range of the technique, 
we note that the scattered intensity for 
acoustic modulations in (123)O$_{7-x}$ is proportional to the in-plane susceptibility 
$\chi^{\prime\prime}(q_x,q_y,\omega)$ \cite{tranquada,note1}
$$I({\bf q},\omega) \propto {k_f\over k_i} |f_{\rm Cu}({\bf q})|^2 
\sin ^2({1\over 2}\Delta z q_z) [n(\omega)+1]\chi^{\prime\prime}(q_x,q_y,\omega) ,
$$
where $k_i$ and $k_f$ are the initial and final neutron wave numbers, $f_{\rm Cu}({\bf q})$
is the Cu$^{2+}$ magnetic form factor, $\Delta z$ (=3.342 \AA) the separation of the 
CuO$_2$ bilayers, $\bf q$ the total momentum transfer 
($|{\bf q}|^2=q_x^2+q_y^2+q_z^2$), and $[n(\omega)+1]$ the Bose population factor.
The solid line in Fig. 1(b) shows the calculated 
$I({\bf q},\omega)$ at $(\pi,\pi)$ as a function of energy 
transfer (along $q_z$) assuming $\chi^{\prime\prime}(q_x,q_y,\omega) =
F(q_x,q_y)\chi^{\prime\prime}(\omega)\propto \omega F(q_x,q_y)$ 
\cite{note2}. Although there are two broad 
peaks in the figure, the observed intensity will  
mostly stem from fluctuations around the lower energy one ($10<\Delta {\rm E}<30$ meV) 
because  of 
the decreased resolution volume at large energy transfers. 
Since room temperature
 triple-axis measurements show 
no detectable magnetic peaks at $(\pi,\pi)$ below 40 meV  (see Figs. 2 and 3),
we have used the integrated scan at 295 K as the background and assumed subsequent net 
intensity gains above the multi-phonon background at lower temperatures are magnetic in origin. 
Figure 1(c) shows
the result at different temperatures. At 200 K, 
the magnetic fluctuations are 
broadly peaked at $(\pi,\pi)$. On cooling to 
150 K and 100 K, the peak narrows in width and grows in intensity but is still 
well described by a single Gaussian centered at $(\pi,\pi)$. 
At 65 K, the data show a flattish top 
similar to previous observations \cite{tranquada}. 
Although detailed 
analysis suggests that the profile is better described 
by a pair of peaks
(Lorentzian or Lorentzian-squared line-shape) 
than a single Gaussian, the most drastic change in the profile 
comes in the low temperature 
superconducting state.  
Rather than the expected single peak, 
two peaks at positions displaced by $\pm\delta$ ($0.057\pm0.006$ r.l.u.) 
from $H=0.5$ are observed, accompanied by a drop in
the spin fluctuations at the commensurate position.
 The observation of sharp incommensurate peaks with the filter integration
technique suggests that the incommensuration 
must be weakly energy dependent in the integration range.

Although the integration technique is excellent in finding small peaks from the 
scattering of lower dimensional objects,
it is important to confirm the result with conventional triple-axis measurements 
and to determine the symmetry of the incommensuration. 
For this purpose, we have realigned the sample in the $(H,3H,L)$ zone. 
If the 15 K profile in Fig. 1(c) stems from an incommensurate structure 
with peaks at $(0.5\pm\delta,0.5\pm\delta)$ [see Fig. 1(a)], 
scans along the $[H,3H]$ direction are expected to peak at $H=0.477$ and 0.523 r.l.u. 
for $\delta=0.057$. 
On the other hand, if the underlying symmetry is
identical to that of 214 [rotated 
45$^\circ$ from Fig. 1(a)], the incommensuration in a $[H,3H]$ scan
 should occur at $H=0.466$ and 0.534 r.l.u.  
Figure 2 summarizes the result at 24 meV \cite{note3}. The scattering 
 at room temperature shows no well defined broad peak around $(\pi,\pi)$,
 but at 70 K a two peak structure emerges.
On cooling below $T_c$, the spectrum rearranges itself with a suppression of 
fluctuations at commensurate point accompanied by an increase in intensity at 
incommensurate positions. The wave vectors of the peaks in the
 $[H,3H]$ scan are consistent with incommensuration at $(0.5\pm\delta,0.5\pm\delta)$. 
 It may also be possible to interpret the data with other structures, however,
we will assume the symmetry of Fig. 1(a)
 until more precise measurements are made.

In previous work, 
superconductivity was found to induce a strong enhancement in the  
$\chi^{\prime\prime}({\bf q},\omega)$ at $(\pi,\pi)$ for
ideally \cite{mignod,mook,fong,bourges} and underdoped \cite{dai1,fong1} (123)O$_{7-x}$ at the resonance positions. 
 Although the intensity gain of the resonance below $T_c$ is 
accompanied by a suppression of fluctuations at frequencies above it  
for the underdoped compounds \cite{dai1,fong1},
no constant-energy scan data are available
 at energies above the resonance. 
In light of the present result at 24 meV for the (123)O$_{6.6}$ sample which
 has a resonance at 34 meV \cite{dai1}, it is important to collect data at these frequencies. 
Thus, we undertook additional measurements with improved resolution (collimation of 
50$^{\prime\prime}$-40$^{\prime\prime}$-40$^{\prime\prime}$-120$^{\prime\prime}$ )
 in the hope of resolving possible 
incommensuration at high energies. Figures 3(a) and (b) suggest that the 
 fluctuations at the resonance energy are  
commensurate above and below $T_c$ with no appreciable change 
in width. For an energy above the resonance (42 meV), the 
scan is featureless at room temperature but shows a
well defined peak centered at $(\pi,\pi)$ at 75 K. Although superconductivity suppresses the 
magnetic fluctuations [see inset of Fig. 3(d)], the wave vector dependence of the line-shape cannot
be conclusively determined due to the poor instrumental resolution at this 
energy.
Unfortunately, further reduction in resolution volume is impractical due to a concomitant drop in the scattering intensities.

Since the earlier polarized work \cite{dai1} has shown that 
for (123)O$_{6.6}$ the 34 meV resonance is the 
dominante feature of $\chi^{\prime\prime}({\bf q},\omega)$ at $(\pi,\pi)$ in the low 
temperature superconducting state, it is important to compare the 
newly observed incommensurate peaks to the intensity gain 
of the resonance. Figure 4 shows the difference spectra between 15 K and 75 K 
at frequencies below and above the resonance. 
In the energy and temperature range of interest (15 K to 75 K),
the phonon scattering changes negligibly  
 and the Bose population factor $[n(\omega)+1]$ modifies the scattered intensity 
at high temperatures by only 
3\% at 24 meV and less at higher 
energies. Therefore, the difference spectra in the figure can be simply regarded as 
changes in the dynamical susceptibility. Inspection of Figs. 4(a) and (b) reveals that the 
susceptibility at 
the incommensurate positions increases on cooling from the normal to the superconducting
state, accompanied by a suppression of fluctuations at the commensurate point.  
 Comparison of Fig. 4(c) to 
 Figs. 4(a) and (b) indicates that 
 the net gain in intensity at the incommensurate 
positions below $T_c$ is much less than that of 
the resonance. For an energy transfer of 42 meV,
the intensity drop appears
 uniform throughout the measured profile, however, instrumental
resolution may mask any possible incommensurate features.
Figure 4(e) plots a summary of the triple-axis measurement in 
superconducting state. Although there are only two constant-energy scans for 
frequencies below the resonance, these data nevertheless confirm the 
result of the integrated technique.

In conclusion, we have found that the low frequency magnetic fluctuations in 
a (123)O$_{6.6}$ sample are incommensurate at low temperatures with 
incommensuration first appearing at temperatures above $T_c$. The
dynamic susceptibility at incommensurate positions increases on cooling below
$T_c$, accompanied by a suppression of magnetic fluctuations at the commensurate 
point.

We thank G. Aeppli, V. J. Emery, K. Levin, and D. Pines for helpful discussions. 
We have also benefited from fruitful interactions with J. A. Fernandez-Baca, R. M. Moon,
S. E. Nagler, and D. A. Tennant.
This research was supported by the US DOE under 
Contract No. DE-AC05-96OR22464 with Lockheed Martin 
Energy Research Corp.

\begin{figure}
\caption{(a) Diagram of reciprocal space probed in the experiment.
The dashed arrow indicates the scan direction with the integrated 
technique while the solid arrow represents the triple-axis 
measurements. (b) Calculated scattered
intensity $I({\bf q},\omega)$ as a function of energy transfer. The effective energy
integration range is mostly from 10 to 30 meV. Note $k_f$ is parallel to $q_z$ in this
technique.
 (c) Integrated measurements in which the data at 295 K are subtracted from 
200 K, 150 K, 100 K, 65 K, and 15 K. The data are normalized 
to the same monitor count. The solid lines in
the 100 K, 150 K, and 200 K data are fits to single Gaussians and linear backgrounds. 
 The solid lines in 
the 15 and 65 K data are two Lorentzian-squared peaks on linear backgrounds which
best fit the data.
}
\label{autonum}
\end{figure}

\begin{figure}
\caption{Triple-axis scans along $(H,3H,1.7)$ at 
24 meV for (a) 295 K,  (b) 70 K,
(c) 58 K, and (d) 50 K.  Data at 295 K were
collected with HB-1 while other scans were taken
using HB-3. Since room temperature measurements at 42 meV [Fig. 3(c)] show no peak 
at $(\pi,\pi)$, the weak structures in (a) are most likely due to 
phonon and/or spurious processes. The horizontal bar shows the
resolution along the scan direction and the vertical
resolution is 0.14 \AA$^{-1}$.
The positions of incommensuration at $H\approx0.48$ and 0.53 r.l.u. are 
indicated by the arrows.
Solid lines in (b)-(d) are two Lorentzian-squared  peaks
on a linear background. The increased scattering at $H>0.6$ r.l.u. 
is due to phonons.
}
\end{figure}

\begin{figure}
\caption{Constant-energy scans along $(H,3H,1.7)$ with energy 
transfer of 34 meV at (a) 75 K, and (b) 15 K.  
Identical scans at 42 meV at (c) 295 K ($\bullet$), 75 K ($\circ$), and (d) 15 K ($\circ$). 
Inset ($\bullet$) shows
the temperature dependence of the scattering at (0.5,1.5,-1.7) for $\Delta E=42$ meV 
where the arrow indicates $T_c$.
The multi-phonon background in the 295 K data has been scaled to the  
value at 75 K for clarity.  Solid lines are Gaussian fits to the data.
 The ${\bf q}$ width of the resonance after deconvolving the 
resolution (horizontal bar) is 0.23 \AA$^{-1}$.  
A similar result is obtained at 42 meV.}
\end{figure}

\begin{figure}
\caption{Difference spectra along $(H,3H,1.7)$ between low temperature ($<T_c$) and
high temperature ($\approx T_c+12$ K) at (a) 24 meV, (b) 27 meV, 
(c) 34 meV, and (d) 42 meV. All data were taken with the same 
monitor units. To a very good approximation, the data can be regarded
as the difference in susceptibility between 15 K and 75 K, 
{\it i.e.}, $\chi^{\prime\prime}(15$ K$)-\chi^{\prime\prime}(75$ K). Solid
lines are guides to the eye. (e)
Summary of triple-axis measurements. Open squares indicate incommensurate positions.
Solid and open circles are the resonance and fluctuations at 42 meV, respectively.
The error bars show the energy resolution and the intrinsic ${\bf q}$ width (FWHM).}
\end{figure}


\begin{references}
\bibitem{note2} Here we have adopted the notation of Ref. \cite{tranquada}. 
The expression $\chi^{\prime\prime}(\omega) \propto\omega$ is valid 
for $\omega\rightarrow 0$ \cite{zha1} and we have extended it to give an 
 estimate of the integration range.
\bibitem{note3} In Ref. \cite{dai1}, we
have observed a drop in magnetic signal at $(\pi,\pi)$. 
Unfortunately, the polarized constant-energy scan
at 24 meV (see Fig. 2(b) of Ref.\cite{dai1}) did not pick up the
incommensurate feature described in this work due to limited 
intensity of the technique.
\end{references}
\end{document}